# Development of a measurement system enabling the reconstruction of γ-ray time spectra by simultaneous recording of energy and time information

Hiroyuki Tajima[1,2]*, Shinji Kitao[1], Ryo Masuda[1], Yasuhiro Kobayashi[1], Takahiko Masuda[3], Koji Yoshimura[3], and Makoto Seto[1]*

[1]*Institute for Integrated Radiation and Nuclear Science, Kyoto University, Kumatori, Osaka, 590-0494, Japan*

[2]*Division of Physics and Astronomy, Graduate School of Science, Kyoto University, Kitashirakawa, Sakyo, Kyoto 606-8502, Japan*

[3]*Research Institute for Interdisciplinary Science, Okayama University, 3-1-1 Tsushima-naka, Kita-ku, Okayama 700-8530, Japan*

We developed a measurement system that enables the reconstruction of γ-ray time spectra in cascade decay schemes. As this system records all the time and energy information of γ-rays, reconstruction is possible after the measurement. Therefore, the energy regions for the γ-ray identification can be optimized in complicated cascade decay schemes. Moreover, in this system we can record data with time-dependent parameters of external perturbations, such as applied magnetic fields, and consequently we can investigate the correlations and responses of γ-rays to the perturbation. This property fulfills the demands required for quantum information research with γ-rays.

The γ-rays emitted from radioactive isotopes (RI) have various useful properties. Because of the diverse characters of nuclides, various combinations of the energy $E$ and the energy width $\Delta E$ of γ-rays are available. For example, Mössbauer spectroscopy, which is one of the most effective methods for condensed matter science, uses γ-rays with a high energy resolution that typically reaches $\Delta E/E \sim 10^{-13}$, which can well resolve the hyperfine interaction in the nuclear levels through atomic electrons[1]. On the other hand, the fundamental processes and control of γ radiation have been extensively studied [2, 3]. Recently, γ-ray photons have attracted much attention in studies on quantum information and communication because γ-ray photons have the following advantageous properties compared with optical photons: high spatial resolution, high transparency in many materials, and in some cases ultra-high energy resolution. In recent research, coherent control of the waveforms of recoilless γ-ray photons using a cascade scheme of the decay of Mössbauer sources was reported[4]. In the cascade decay scheme using Mössbauer effects, a heralding γ-ray photon is emitted when a nucleus decays from the higher excited state to the lower intermediate state, and a consequent γ-ray photon whose waveform is controlled is emitted with a transition from the intermediate state to the ground state. The heralding γ-ray photon is used as a reference start signal for the measurement and control of the waveform of the consequent photon in the time domain. In these measurements, the photons are identified by their energies to observe the time correlation between the first γ-ray photon and the successive intended photon, as unnecessary emissions such as γ-rays by other transitions and fluorescent X-rays are also observed. Their energy is usually identified with electronic apparatus such as single channel analyzers (SCA), which select signals with specific pulse heights. After the selection of signals by SCA, the time spectra are measured using either time to amplitude converters (TAC) with multi-channel analyzers (MCA) or time to digital converters (TDC). In this approach, the energy regions of the γ-ray photons should be determined in advance. However, it is sometimes difficult to determine the energy region of interest in complicated decay schemes. Moreover, it is necessary to understand the time and energy of each detected γ-ray photon to observe the response of the γ-ray waveform to external perturbations, such as time dependent magnetic fields and energy modulations applied to the γ-ray sources or absorbers. Therefore, it is required to develop a flexible measurement system that fulfils these demands to advance quantum information research using Mössbauer γ-ray photons. Recently, a system for time spectrum measurement with a high speed multi-channel scaler (MCS) has been developed[5], which has been used for time spectrum measurement of the 26.27-keV level of $^{201}$Hg excited by synchrotron radiation[6]. In this research, we constructed a MCS system that records information of the energy and time of each detected γ-ray photon simultaneously. This system allows us to reconstruct the waveform of the intended γ-ray photons by selecting the energy after measurement from the recorded data. Using this system, we demonstrated the ability to observe the γ-ray decay time spectra using a cascade decay scheme.

As the excited states of $^{181}$Ta have been used in time differential perturbed angular correlation (TDPAC) measurement, they are suitable to confirm the assurance of the constructed system in this study. The excited states of $^{181}$Ta were created as β-decay products of $^{181}$Hf that were generated by irradiating $^{180}$Hf in a natural Hf foil (15 × 15 × 0.025 mm$^3$) in the pneumatic tube station Pn-2 of the Kyoto University Research Reactor (KUR) in the 1MW operation mode with the thermal neutron flux of 5.5 × 10$^{12}$ neutrons/(s·cm$^2$) for 1 hour. After the β decay process, the eleventh excited state of $^{181}$Ta emits 133 keV γ-rays decaying to the sixth excited state and then emits 482 keV γ-rays decaying to the ground state with a half-life of 10.8 ns. We measured the time spectra of the 482 keV γ-rays while the 133 keV γ-rays were used as a starting reference. BaF$_2$ scintillators were used as γ-ray detectors because they have a high time resolution ($\Delta t \simeq 300$ ps). In our measurement, we used two BaF$_2$ scintillators denoted as "BaF$_2$_1" and "BaF$_2$_2" in this note. Each signal is amplified with a fast preamplifier (Phillips Scientific 6954). A signal from each detector is divided into two signals; one is input to a constant fraction discriminator (CFD), which corrects the time shift due to the pulse height difference of the input analog signals and outputs a Nuclear Instrumentation Module (NIM) logic signal as a timing signal. The other signal is input to an amplitude to time converter (ATC), which outputs a logic signal with a delay proportional to the pulse height of the input analog signal.[5]. We used a MCS, which measures absolute time for five input channels with 100 ps time resolution (MCS6, FAST ComTec). The CFD output signals are recorded in the MCS as timing information. Moreover, the energy information is also recorded in the MCS as timing information converted by the ATC; by using the CFD output signal as a reference, the time difference between the CFD signal and the delayed ATC output signal corresponds to the incident γ-ray energy. In the ATC module, CFD output signals are also used to activate the ATC function (Fig.1). In our measurement, the CFD outputs were input to Ch. 1 and Ch. 3 of MCS, corresponding to BaF$_2$_1 and BaF$_2$_2, respectively, and the ATC outputs were input to Ch. 2 and Ch. 4, also corresponding to BaF$_2$_1 and BaF$_2$_2, respectively. Therefore, the time difference between Ch. 1 and Ch. 2 expresses the energy of the γ-ray photon entering BaF$_2$_1. In the same way, the time difference between Ch. 3 and Ch. 4 expresses that entering BaF$_2$_2. After the end of the measurement, using these relationships the energy and absolute time information of all incident γ-ray photons can be obtained.

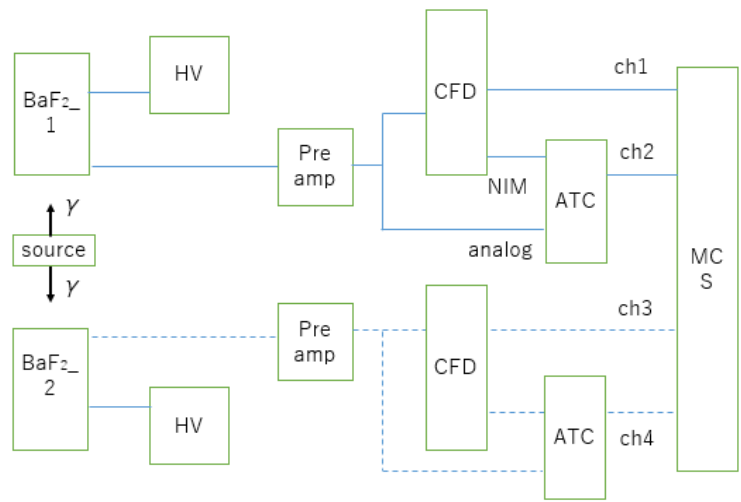

**Fig. 1** Block diagram of the system for simultaneous time and energy measurement (ATC: amplitude-to-time convertor; MCS: Multi channel scaler; CFD: constant fraction discriminator; HV: high voltage).

The constructed energy spectrum of γ-ray photons detected by BaF$_2$_1 using the time difference between Ch. 1 and Ch. 2 is shown in Fig. 2(a); for comparison, the energy spectrum obtained by a usual MCA is also shown in Fig.2(b). In the MCA spectrum, the second peak from the left is due to 133 keV γ-rays and the fourth peak to 482 keV. There are also 346keV γ-rays unnecessary for our measurement as the third peak. In the energy spectrum obtained with the ATC and MCS shown in Fig. 2(a), the horizontal axis unit is written in nano-seconds (ns) because the pulse height information is transformed into time by the ATC. As the ATC response to input signals with lower pulse height (under 200mV) becomes worse, the threshold of the CFD enabling the output signals is set somewhat high (250mV). Therefore, signals with lower pulse heights under the value corresponding to 70 ns are not observed in the ATC spectrum. In Fig. 2(a), the left sharp peak and the right broad peak are due to 133 keV γ-rays and 482 keV γ-rays, respectively. However, as the resolution of the ATC in the high energy range is worse than the MCA measurement, the broad peak around 230 ns (400 keV) is due to unresolved 346 keV and 482 keV γ-rays. The relation between the energy and the ATC delay time indicated as the upside horizontal axis and the downside one in Fig. 2(a) was evaluated by least square fitting with Gaussian profiles referencing the intensity ratio in the MCA spectrum. However, as mentioned above, as the ATC response is worse in the low energy range, the 133 keV peak position was estimated by considering the intensity (area) ratio of the 133 keV to the sum of the 346 keV and 482 keV peaks in the MCA spectrum. Further, 159 keV γ-rays from $^{123m}$Te and 364 keV γ-rays from $^{131}$Xe that are generated from $^{131}$Te through $^{131}$I, which were emitted from a neutron irradiated natural Te source, were used for the energy calibrations. Using these assignments of the ATC time delay spectrum to the γ-ray energy, the discerned signals corresponding to start signals (133 keV γ-ray photons) and stop signals (482 keV γ-ray photons) enable the construction of the decay time spectrum; the 90–140 ns region in Fig. 2(a) was assigned as 133 keV start signals and the 200–400 ns region was assigned as 482 keV stop signals for both BaF$_2$_1 and BaF$_2$_2. When two correlated γ-ray photons radiated in a cascade decay series enter the different detectors, they are recoded as timing input signals to Ch. 1 and Ch. 3. Although they are assigned by their energies, uncorrelated γ-ray photons sometimes enter accidentally. Therefore, in our analysis the correlation time was set as 100 ns because the probability of counting correlated signals outside of the correlation time is very low due to the expected half-life of around 10 ns and the count rate of about 6000 counts per second. Therefore, we limited the time spectra to within 100 ns. With these evaluations of the signals, the decay time spectrum can be constructed. It is noted that, if a start 133 keV γ-ray photon is detected by one of two detectors in this system the corresponding stop 482 keV γ-ray photon is expected to be detected by the other detector, because the total dead time of the detector and ATC is estimated as about 400 ns, which is longer than the expected half-life of about 10 ns. Therefore, we can construct two time spectra by selecting the combinations (BaF$_2$_1:start - BaF$_2$_2:stop, BaF$_2$_2:start - BaF$_2$_1:stop), and these were added. Measured time spectrum constructed in this way is shown in Fig.3. The observed background of about 180 counts is caused by the signals of uncorrelated gamma photons due to chance coincidence. Assuming exponential decay, least-squares fit was performed to the observed spectrum and a half-life of 10.6±0.3 ns was obtained. This value agrees well with the previously reported value of 10.8±0.1 ns within the experimental error[7]. This result shows that the constructed system works well for the time spectrum measurement of γ-rays emitted from nuclides with cascade decay schemes.

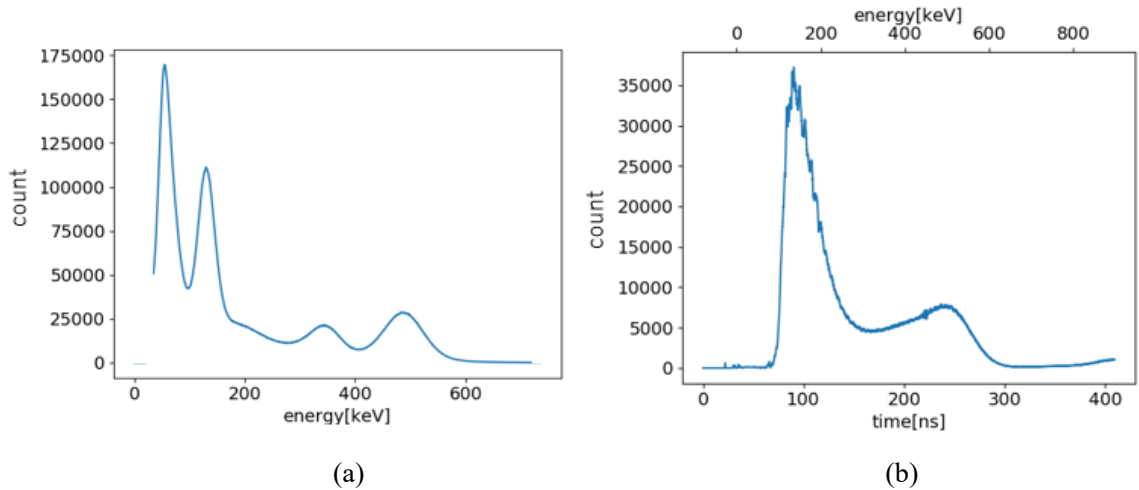

(a)          (b)

**Fig. 2** Measured energy spectra of $^{181}$Hf decay with the ATC and MCS (a) and with a conventional MCA (b).

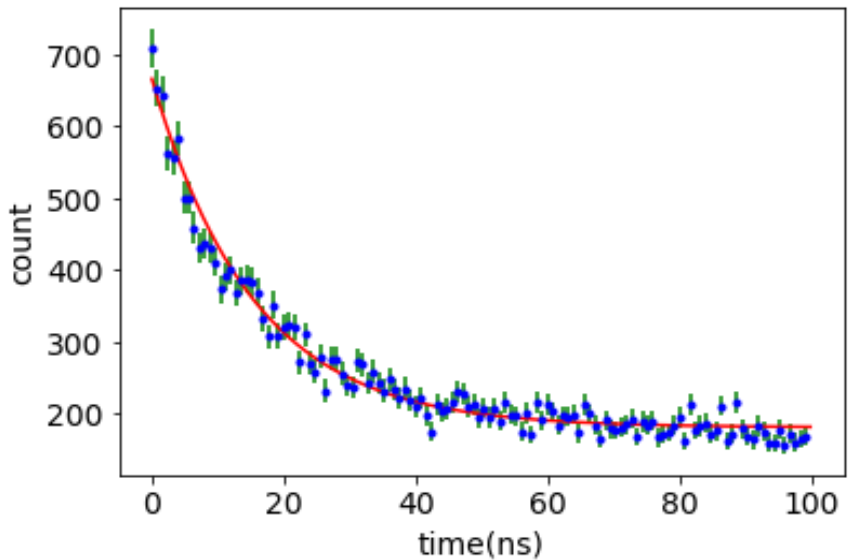

**Fig. 3** Time spectrum constructed from the data measured by a simultaneous time and energy measurement system.

In this method it is easy to modify the γ-ray energy regions of interest after the measurement and therefore we can optimize the signal-to-noise (S/N) ratio of the spectrum. In contrast, in conventional methods using SCA, as the energy regions of interest are fixed before the measurement this type of reconstruction is impossible afterward and the optimization of the region of interest may require re-measurement even if it takes a long time. In fact, the decrease of the radiation intensity of RI sometimes demands the resetting of the regions even during a measurement. Therefore, this system may be useful for TDPAC measurement. Furthermore, time dependent perturbations such as modulation of external magnetic fields, and oscillation of the source and/or absorber may be required for the waveform control of γ-ray photons[4]. Using our newly developed system, we can extract the response of the waveforms of the γ-rays to these external perturbations in specific parts of interest after the measurement by inputting the parameter signal of the perturbation to the MCS. Therefore, the developed system fulfills these measurements to advance quantum information and communication research using Mössbauer γ-ray photons. However, one disadvantage of this system is that it requires a large amount of data storage. In this method of measurement, typically 4 GB data are stored for one spectrum. However, the recent development of large-capacity storage devices seems to overcome this problem. Although the MCS used in this measurement enables five channel inputs, more than one MCS can be used with common synchro-signals, which allows for the response measurement of many parameters and detectors. Hence, our developed system will improve not only the S/N ratio but also extend the degrees of freedom of the measurement.

**Acknowledgments**

We thank Dr. M. Saito and Dr. M. Kurokuzu at the Institute for Integrated Radiation and Nuclear Science, Kyoto University for their experimental support and helpful discussion. This work was supported by JSPS Grant-in-Aid for Scientific Research Grant no. 16K13723. This work has been performed by using facilities of the Institute for Integrated Radiation and Nuclear Science, Kyoto University.